\documentclass{aa}
\usepackage{txfonts}
\usepackage[dvips]{graphicx}
\usepackage{natbib}
\usepackage{longtable}
\bibpunct{(}{)}{;}{a}{}{,}

\begin{document}

\title{Validation of the new Hipparcos reduction}

\author{Floor van Leeuwen}
\institute{Institute of Astronomy, Madingley Road, Cambridge, UK}

\date{Received 27 July 2007 / Accepted 8 August 2007 }

\abstract{A new reduction of the astrometric data as produced by the Hipparcos
mission has been published, claiming accuracies for nearly all stars brighter 
than magnitude $\mathrm{Hp}=8$ to be better, by up to a factor 4,
than in the original catalogue.}{The new Hipparcos astrometric catalogue
is checked for the quality of the data and the consistency of the formal 
errors as well as the possible presence of error correlations. The differences 
with the earlier publication are explained.}{The internal errors are followed 
through the reduction process, and the external errors are investigated on 
the basis of a comparison with radio observations of a small selection of 
stars, and the distribution of negative parallaxes. Error correlation levels 
are investigated and the reduction by more than a factor 10 as obtained in
the new catalogue is explained.}
{The formal errors on the parallaxes for the new catalogue are confirmed. 
The presence of a small amount of additional noise, though unlikely, 
cannot be ruled out.}
{The new reduction of the Hipparcos astrometric data provides an improvement
by a factor 2.2 in the total weight compared to the catalogue published in 
1997, and provides much improved data for a wide range of studies on stellar
luminosities and local galactic kinematics.}


\maketitle

\section{Introduction}

Since the publication of the Hipparcos catalogue \citep{esa97} there have 
been suggestions, though not always well founded, of the presence of 
systematic errors in the astrometric data the catalogue contains 
\citep[see for example:][]{naray99, soder05}. Suspicion was in particular 
raised concerning the 
reliability of the parallaxes at the sub-milli arcsecond level. Such errors, 
if present, may become apparent when for example averaging parallax 
information for open clusters. Here systematic or correlated errors would 
be the most conspicuous and damaging, as the formal accuracy on a cluster 
parallax can well exceed those of individual stellar parallax determinations 
in the catalogue. Some problems with the Hipparcos astrometric data, as
due to inaccuracies in the along-scan attitude reconstruction, have 
since been suggested by \citet{makar02} and identified in detail by 
\citet{fvl05a}. As these problems were found to be 
curable, the concept of a new reduction \citep{fvl05b} became a viable option. 
This reduction was recently completed and a full description of the processes 
used is now in press \citep{fvl07}. In that 
publication, however, the emphasis is on illustrating the proper use of 
the Hipparcos astrometric data and the way these have been obtained in the 
new reduction. The data used to illustrate these processes were obtained from
(mainly nearly-final) iteration phases. The present paper provides the overall
quality check on the new catalogue as it will become available in 2007 through
the data disk included in \citep{fvl07} and early 2008 through AstroGrid
\footnote{http://www2.astrogrid.org/}.

With the availability of a new and potentially more accurate astrometric 
catalogue, it is now also possible to do a detailed investigation of the
catalogue published in 1997. The importance of this investigation is that
any observed differences need to be fully understood from the differences
between the old and new reduction methods.

One may get the impression that errors were made in the original reduction
of the Hipparcos data, but this is not the correct way of describing what 
has happened. Several aspects have to be considered before judging the 
performance of the original reduction by the two consortia, 
FAST \citep{koval92} and NDAC \citep{lindeg92}:
\begin{enumerate}
\item The original aim of the mission was to achieve a parallax accuracy 
of around 2~milli~arcseconds (mas). The final results as published in 1997 were
about a factor two better, despite a very serious problem with the orbit
of the satellite;
\item Computing hardware available around 1993 to 1996, when the final 
calculations took place, was much less powerful than what is available today.
This reflected in the time required for a single iteration
(more than 6 months), and a limitation to how often this could be done. It also
limited the possibility to interact efficiently with the reduced data to
check results. What took more than 6 months some 12 years ago, takes currently
about a week on a single desktop computer;
\item The Hipparcos measurement principle is complex and sensitive. Some of
this was well understood before the start of the mission, not least thanks
to the preparatory work of Lennart Lindegren. An input catalogue 
\citep{turon92} with an even distribution of target stars over the sky 
was created to ensure a stable astrometric solution for the mission data. 
How sensitive the solution really is became only clear during the iterations
of the new reduction. In particular, the principal requirement of linking the 
data in the two fields of view is critical, and links can far more easily be 
weakened than was imagined 25 years ago, when in the 1980s the input 
catalogue was constructed.
\end{enumerate}
This paper, and the two associated with it \citep{fvl05a,fvl05b}, are 
therefore not an accusation of bad workmanship in the creation of the 
catalogue published in 1997. The intention is rather to show that the 
principle of the Hipparcos instrument, now also being implemented for the 
Gaia satellite, is valid and is the only way one may measure directly, 
fully reliable absolute parallaxes, and do so for large numbers of stars. 
These papers show that the reduction of the Hipparcos data is far from 
simple, but also that complications associated with the reduction of those 
data are now much better understood, and can be sufficiently controlled.

For clarity of the current paper, the main issues that have been resolved
in the new reduction, and which have led to a significant improvement in
accuracies for the new catalogue are briefly summarized here. For a more
comprehensive description the reader is referred to the two papers already
mentioned above \citep{fvl05a, fvl05b} and to \citet{fvl07}. 

The noise on the astrometric data as gathered by Hipparcos originates from 
two main sources. The first source is the Poisson noise on the original photon 
counts from which the measurements of transit times were derived, and which 
sets the accuracy limit for those data. The second is the noise from the 
along-scan attitude reconstruction, which provides the reference frame used to
transform transit times to one-dimensional positions on the sky, a 
transformation that needs to be free from distortions. Significant 
improvements in the attitude reconstruction have been obtained in the new 
reduction through a much improved understanding of peculiarities in the 
dynamics of the satellite, in particular concerning
non-rigid events (scan-phase discontinuities) and hits by dust particles. 
Abandoning the great-circle reduction technique \citep{vdmar88,vdmar92} 
for a global iterative solution allowed for a better control over the 
attitude reconstruction, but makes it necessary to iterate the reduction
between the attitude reconstruction and the reconstruction of the astrometric
parameters. A similar iteration is also planned, and has already been tested
extensively \citep{omullane06}, for the Gaia data reductions \citep{lindeg05}.

The final improvement for the new reduction comes from a better understanding
of the connectivity requirement in the reconstruction of the along-scan 
attitude. The connectivity requirement, when applied correctly, enables
sufficient contributions to the along-scan attitude from both fields of 
view of the telescope. This is essential for obtaining a reconstructed sky
free from local small distortions. Such distortions can ultimately lead to
local variations in the parallax zero point. Weights of data contributions from
the two fields of view can in practice fluctuate considerably, despite
the construction of the Hipparcos Input Catalogue, which was designed to 
provide a more or less homogeneous distribution of selected stars over the
sky \citep{esa92}. The weight-ratio allowed in the along-scan attitude
reconstruction for the two fields of view affects the convergence of
the final catalogue as well as the noise level for the brightest stars. 
Experiments done during the construction of the new catalogue arrived 
at an optimal value of around 2.7. A larger value would make the convergence
move along very slowly, a smaller value would create unacceptable noise levels
for the brightest stars. 

The Hipparcos data are obtained in the form of transit times, which are
transformed to angular positions along the scan directions. These positions
are referred to as asbscissae, and in all applications the input
consists of abscissa residuals, the differences between the observed and
predicted angular positions along the scan direction. This definition of the
abscissae is different from that used in the original reductions, where
abscissae referred to angular positions as projected onto a reference great 
circle. In the new reduction, the abscissae are measured strictly with 
respect to the instantaneous scan direction.

The present paper will refer to various observational aspects of the 
Hipparcos mission, without going into detail. Extensive descriptions
on how the Hipparcos data were obtained can be found in Volume~3 of
\citet{esa97}, as well as in \citep{fvl07}. 

This paper has been organized as follows. Section~\ref{sec:internal}
summarizes the statistical tests carried out on the internal formal 
errors at the relevant stages of the data reductions. Also considered
are correlation levels of the abscissa residuals and the global
dependencies of the formal errors on the astrometric data.
Section~\ref{sec:external} presents the external verification of the
data, though the possibilities for this are rather limited at 
accuracies better than 0.3~mas. Section~\ref{sec:impact} gives a brief
impression of the potential impact of the new reduction. Finally, 
Section~\ref{sec:conclusion} presents a summary of the conclusions 
derived from this study. 
 
\section{The internal accuracy verification}
\label{sec:internal}

\subsection{The transit data}

\begin{figure}
\includegraphics[width=9cm]{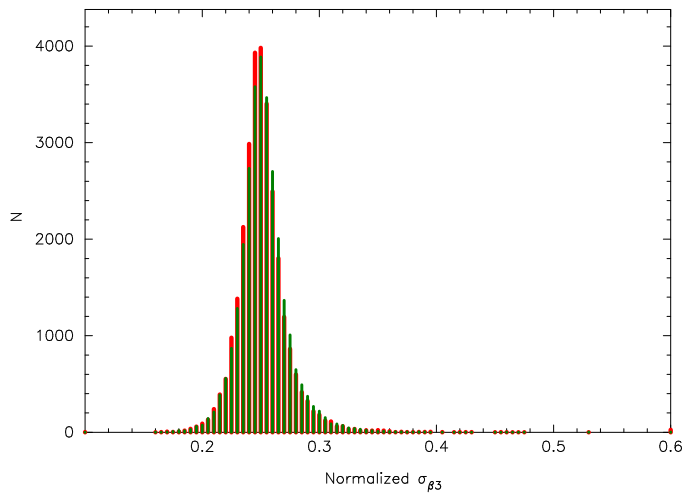}
\caption[]{The normalized errors ($\sigma_{\beta_3}\times{M_1\sqrt{I_\mathrm{tot}}}$) 
on the modulation phases for transits over the modulating grid. The data are 
for orbit 174, 21 January 1990.}
\label{fig:sb3}
\end{figure}
The positional information obtained by Hipparcos is derived from the
modulated signal created by the transit of a stellar image over a regular
grid of transparent lines. These transits are examined over fixed time 
intervals of just over 2~s, and it took 9 such transits to cross the $0\fdg9$
field of view. The grid, with a periodicity of 1.2~arcsec as projected on the 
sky, creates a signal which is dominated by a first and second harmonic. 
The modulation phases define the transit time for a stellar image with respect 
to a reference grid line. The modulation amplitudes of the harmonics are weak 
functions of stellar colour and position on the grid, dependencies which are 
calibrated and used to detect double stars. Data for different
stars are obtained pseudo simultaneously by switching the small sensitive area
of the photomultiplier detector between different stellar images as these move
across the grid. The first formal error to enter the reduction is therefore
the error on the estimate of the modulation phase, represented 
traditionally by $\beta_3$ \citep[for details see][]{fvl07}:
\begin{equation}
\sigma_{\beta_3} \approx \frac{0.25}{M_1\sqrt{I_\mathrm{tot}}}~{\mathrm{arcsec}}.
\label{equ:sigma}
\end{equation}
Here $M_1$ is the relative modulation amplitude of the first harmonic, with
a typical value between 0.6 and 0.75, mainly depending on the colour index 
of the observed star. The integrated intensity $I_\mathrm{tot}$ is obtained 
from the total photon count of the signal. The scaling factor 
0.25 is stable to within 1~per~cent over the mission, and contains amongst 
others the conversion from modulation phase to positional displacements in 
the focal plane. Figure~\ref{fig:sb3} shows a histogram of the ``normalized'' 
formal errors as observed for the transit data in one orbit near the start 
of the mission. 
\begin{figure}
\includegraphics[width=9cm]{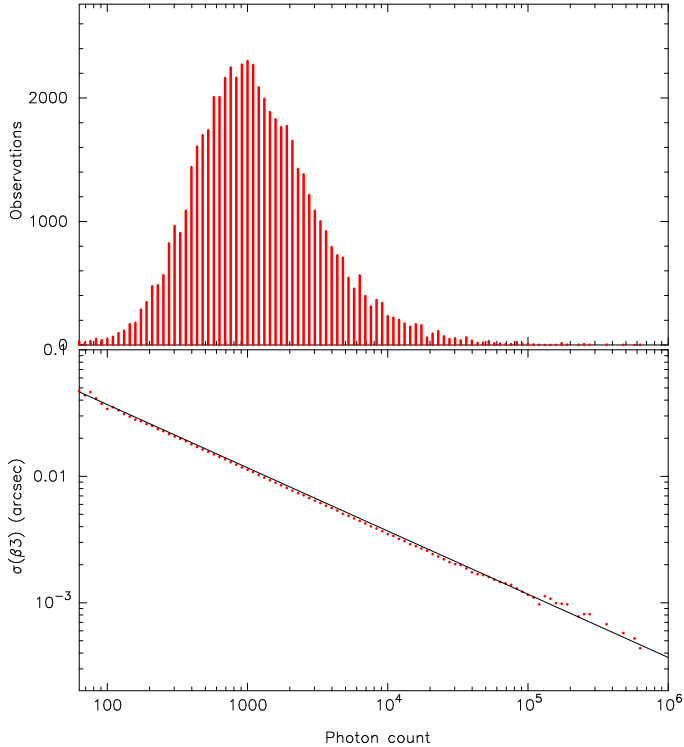}
\caption[]{The distribution of numbers of observations (top) and mean errors
on the modulation phase (bottom), as a function of the integrated photon 
count of the observation. The diagonal line in the lower diagram shows the
slope of the photon noise relation. The offset is fairly arbitrary
because of variations in $M_1$. This also reflects in the variations in the
data for the brightest transits. The data are for orbit 174.}
\label{fig:intens}
\end{figure}

The integrated photon count $I_\mathrm{tot}$ for an observation varied 
considerably, as stars from magnitude -1 to 12.5 were observed. In 
general, the brighter stars were assigned less observing time than the fainter 
ones, but this provided only a weak compensation for the intensity 
differences, as the possible maximum ratio in observing time was only a 
factor 16 (equivalent to 3 magnitudes), while the typical range was more 
like a factor four to six. Figure~\ref{fig:intens} shows an example of the 
distributions of formal errors near the start of the mission. These diagrams 
immediately expose a problem with the published data. The formal errors on 
single transits of the brightest stars are already smaller than the accuracy
of the reconstructed position as derived from the astrometric parameters and 
their covariance matrix in the 1997 publication, which are based on all 
mission data combined. In the old reduction, the noise on the astrometric 
parameters for these stars was entirely caused by inaccuracies in the attitude 
reconstruction, and showed significant correlations for in particular bright 
stars at small separations on the sky (see further 
Section~\ref{sec:correlations}).

\begin{figure}
\includegraphics[width=9cm]{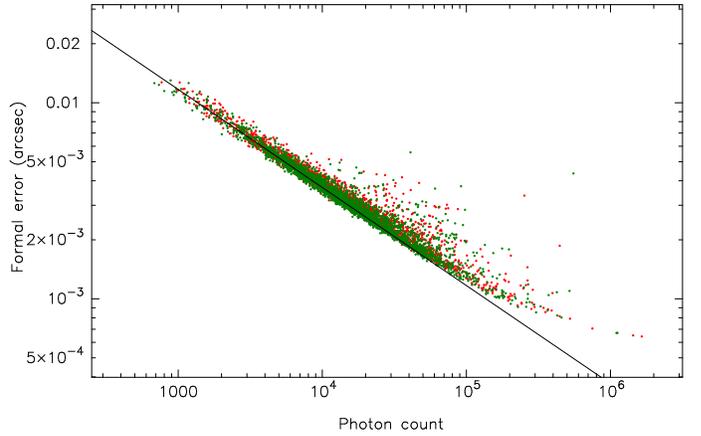}
\caption[]{The formal errors on the combined abscissa residuals, 
including the errors on the predicted positions. The diagonal line is drawn
in the same position as in Fig.~\ref{fig:intens}. The upturn towards
the brighter transits reflects the contribution from the uncertainty in the 
stellar reference positions. The data are for orbit 174.}
\label{fig:attabsc}
\end{figure}
The transit observations are used as input for the along-scan attitude
determination. They are combined with a preliminary estimate of
the attitude (based on star mapper data) and predicted positions from the 
last version of the astrometric catalogue to provide abscissa residuals. 
Studies concerning the dynamics of Hipparcos \citep{paper4} have shown
that regular positional variations with amplitudes above 0.1~mas are 
mostly restricted to time scales above 50~s. It therefore seems justified to
group data over 10~s intervals. There have been noted a few instances when
still not fully-understood disturbances of the rotational motion of the 
satellite took place on shorter time scales. This affected only small parts 
of about half a dozen orbits. For those few data sets, the relevant parameter 
(the density of nodes in the spline fitting) for the attitude model was 
adjusted. The grouping of data, which is done for each field
of view separately, has a number of advantages. It creates pseudo measurements 
with smaller formal errors, which allow for a better detection of attitude
disturbances such as hits and basic-angle drifts. The number of observations 
in an otherwise relatively large solution is reduced by about a factor 5,
requiring less processing time. The formal errors assigned to individual 
(pseudo) measurements take into account the accuracies of the predicted transit
times as derived from the available astrometric solution. This means that 
the formal errors tend to be larger than that given by the photon 
statistics alone, in particular for the brightest stars 
(Fig.~\ref{fig:attabsc}). Where more than one observation of the same star 
contributed to a group, the  contribution of the external error was 
appropriately adjusted.

\subsection{The basic angle and other instrument parameters}

\begin{figure}
\includegraphics[width=7cm]{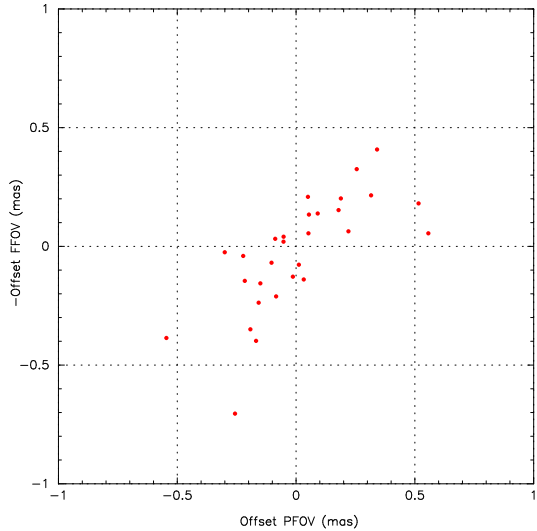}
\caption[]{The correlation between abscissa residuals in the preceding and
following fields of view. The residuals have been averaged over intervals
of 16 minutes (1/8 rotation of the satellite) to increase the resolution. 
The correlation shows small variations of the basic angle taking place at 
time scales longer than 16 minutes. The data are for orbit 174.}
\label{fig:absc_ba}
\end{figure}
Stability of the basic angle is a critical requirement when it comes to
triangularization of the sky with an instrument like Hipparcos. The 
stability of the basic angle is naturally not exact, and some noise 
is left from small variations. The presence of such variations, and
their scale can be demonstrated by means of a detailed study of the residuals
left after the along-scan attitude reconstruction. To make this visible, 
residuals have been averaged over 16 minute intervals (equivalent to
a 45 degrees rotation of the satellite), so that the noise level is reduced 
by nearly a factor 10.

Figure~\ref{fig:absc_ba} is an example of such an abscissa-residual test,
showing a correlation with an amplitude of about 0.4 mas, which would
contribute an additional abscissa noise of 0.3~mas. A test involving 
10 successive well-covered orbits shows the same approximate level 
of the correlations, but there are no signs of systematic variations,
i.e.\ the variations are not linked to the rotation phase or the
orbital phase of the satellite. The observed variations are likely to be 
a reflection of the temperature-control cycle of the payload. As was 
explained by \citet{fvl05a}, a spin-synchronous modulation could have 
led to a systematic offset in the parallaxes, but is clearly not observed.
That such a modulation is unlikely to be present was also shown indirectly 
through the determination of the parallax zero point, which was found to be 
correct to within 0.1~mas \citep{areno95,llind95}. 

A few actual basic-angle drifts have been observed over the
mission, and these are linked to known thermal-control problems of
the payload. A detailed description and identification of the orbits
affected is found in \citet{fvl07}. 

Next to the basic angle, three sets of instrument parameters have been 
used to describe the large-, medium- and small-scale distortions of the 
description of the geometric projection of the sky on the focal plane and the 
measuring grid. The large-scale distortions describe the projection effects of 
the telescope optics and their variation with time. These distortions are 
represented by a two-dimensional third-order polynomial in position and a 
linear colour correction.
The medium-scale distortion corrections reflect the small-scale optical and
the grid distortions, which are observed to be very stable over the mission, 
with a maximum amplitude just below 1~mas. The large- and medium-scale 
distortions are both resolved over the grid. The small-scale distortions 
represent the printing characteristics of the modulating grid, and are 
reconstructed to an accuracy level of approximately 0.1~mas. This correction, 
too, is very stable over the mission. It is resolved only as a function of 
the transit ordinate by means of collecting abscissa residuals left after the 
astrometric solutions, at a resolution of 1000 intervals across scan. 
There is also a fourth correction, which describes the detailed colour 
dependence of the abscissa residuals. The detailed colour and the small-scale 
geometric correction also provide information on local abscissa noise, from 
which formal error corrections have been derived as a function of
transit ordinate and star colour index. These corrections compensate for
inaccuracies in the calculation of the modulation factor $M_1$ in 
Eq.~\ref{equ:sigma}, when calculating formal errors. Further details on these 
calibrations are provided by \citet{fvl07}. For the assessment of the 
internal accuracies of the Hipparcos data, the most important aspect of 
these calibrations is that they appear to have been able, together, to 
represent the geometric distortions down to a level of about 0.1~mas, which
is insignificant with respect to other noise contributions.

\subsection{The along-scan attitude}
The along-scan attitude has been derived using the fully-dynamic model (FDM) as
presented first by \citet{fvl05b}. For details the reader is referred to
that paper or to \citet{fvl07}. Here only a summary relevant to the 
accuracy assessment is presented.

Through the FDM the underlying torques acting on the satellite are 
reconstructed. From these torques are derived the satellite rotation rates 
by means of integration over the Euler equation for the motion of a rigid body 
in space. Error angles are then obtained through integration of the 
rotation rates. Both integrations require starting points, which are also
part of the attitude modelling.
In an iterative fitting procedure for the along-scan attitude, abscissa 
residuals are fitted as a function of time with a 5th order exact spline 
function \citep{paper4, fvl07}, the second derivative of which provides the 
correction to the 
torque model. The integrations are carried out over uninterrupted intervals. 
Interruptions can be caused by thruster firings, scan-phase jumps, 
(micro meteoroid) hits or 
large gaps in the data stream (due to occultations, perigee passages, 
no ground-station coverage etc.). The underlying, continuous, torque model 
covers time intervals between any two gaps in the data that are longer than 
1 minute. For each integration interval, the starting values for the pointing 
and rate integrations are also determined. Thus, the reconstructed attitude 
can be evaluated at any point of time, except for times very close to the 
discontinuities.

The attitude-reconstruction process ensures the proper level of connectivity 
between observations in the two fields of view. This is achieved through
controlling the relative weights of data from the two fields of view 
contributing to the attitude reconstructions. These weights are checked 
for each interval between nodes of the spline function. A maximum
weight ratio of 2.72 is allowed. For any larger ratio, the largest weight
is reduced to 2.72 times the smaller weight for the data contained in
the node interval. Very few exceptions were made to this rule, to allow some
short stretches of time with data from only one field of view.

\subsection{Field-of-view transits and astrometric parameters}
\begin{figure}
\includegraphics[width=7cm]{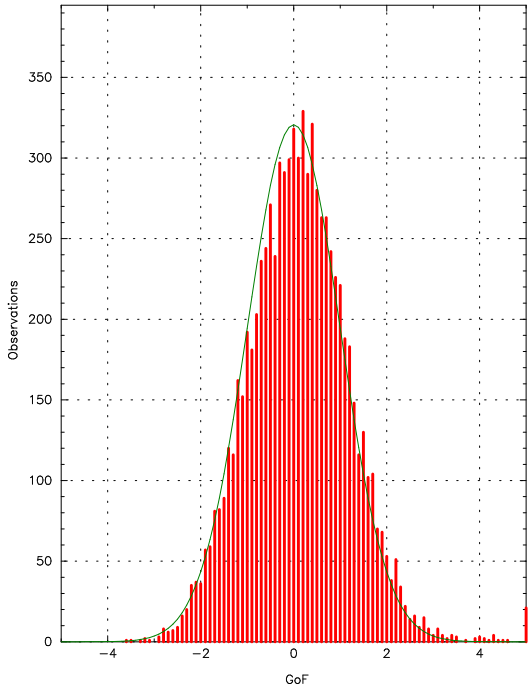}
\caption[]{The goodness-of-fit statistics for the combination of the 
transit data to field transits. The histogram shows the observed distribution,
and the curve the equivalent Gaussian distribution for the same number of 
observations. The data are for orbit 174.}
\label{fig:field_gof}
\end{figure}
The next step in the processing of the transits is the combination of the
abscissa residuals (after correcting for the large- and medium-scale
distortions and determining the along-scan attitude) to mean abscissa
residuals for field-of-view transits. This may involve up to 10 individual 
measurements. A goodness-of-fit statistic is calculated for each field-of-view
transit, based on the formal errors of the transit measurements as described
above. Figure~\ref{fig:field_gof} shows a typical example of the observed
distributions for this statistic. There is an indication of a small 
underestimate of the formal errors on the transit data in this particular
case, shown by the offset of the data from the ideal Gaussian distribution. 
Data of this kind have been collected for all orbits and examined 
for deviations as part of the quality assurance process.

\begin{figure}
\includegraphics[width=9cm]{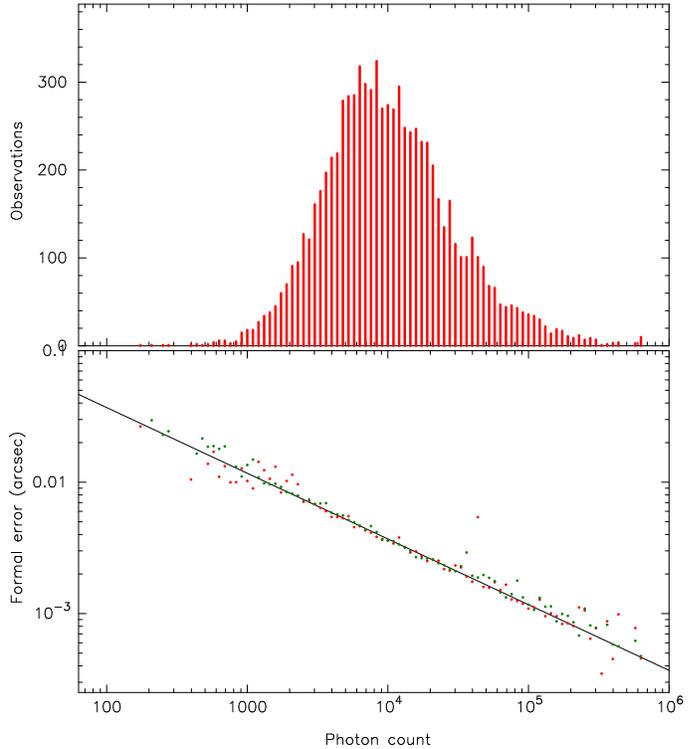}
\caption[]{The distribution of formal errors (bottom) and numbers of 
observations (top) for the field transits. The position of the diagonal line 
in the bottom diagram is the same as in the similar Fig.~\ref{fig:intens} and 
Fig~\ref{fig:attabsc} above.} 
\label{fig:field_fe}
\end{figure}
The formal errors on the field transits are still governed by the same 
dependencies as for the transit data, and are dominated by photon count
statistics, as can be observed in Fig.~\ref{fig:field_fe}. These field
transit data are the input to the astrometric-parameter estimation and the
calibration of the small-scale geometric distortions. They are referred
to as field-transit abscissa residuals (FTAR).

\begin{figure}
\includegraphics[width=9cm]{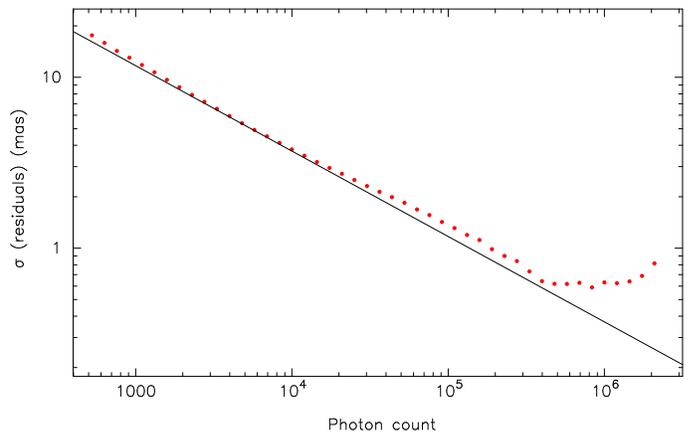}
\caption[]{The dispersions of abscissa residuals after fitting of astrometric
parameters for single stars with five-parameter solutions. The diagonal
line is the same as in the preceding figures, and shows that these residuals
are still largely dominated by photon noise. Only at the brightest end, the
attitude noise (at about 0.6 mas) dominates.}
\label{fig:absc_disp}
\end{figure}
At each cycle of the iterative process that ultimately builds the final 
astrometric catalogue, corrections to the assumed astrometric parameters
have been determined based on the FTARs. These corrections have been 
accumulated as corrections to the astrometry in the catalogue of 1997.
A basic five-parameter astrometric solution has been sufficient for
application to 102~072 out of a total of 117~955 stars. Indicative of the
improvement in quality of the new solution is the number of stars for
which the old catalogue gave a so-called stochastic solution (1561), that
are solved as normal solutions in the new reduction (962). Stochastic 
solutions represent cases where remaining residuals are significantly 
larger than expected. This can be caused by unresolved orbital motion, 
but in case of the published data, also by an accidental accumulation of
unresolved problems in the satellite attitude reconstruction. 

Examination of the dispersions of the FTARs after the astrometric parameter 
solutions (Fig.~\ref{fig:absc_disp}) provides the final check on the
internal consistency of the data, as well as a measure of the remaining
attitude noise. This noise is found to be approximately 0.6~mas as applicable
to a FTAR. In comparison, the attitude noise left in the 1997
catalogue was 1.5 to 2~mas at the level of the combined FTARs for an orbit. 
With on average 4.5 FTARs per orbit, the equivalent attitude noise in the 
original reduction was about 3 to 4~mas, at least a factor five larger than
in the new reduction. This reduction in attitude noise reflects in the 
formal errors of the astrometric parameters, which are determined by 
photon statistics down to about magnitude 3.5 to 4 in the new reduction,
but only down to magnitude 8 to 9 in the 1997 catalogue. 

\subsection{Abscissa-residual error correlations}
\label{sec:correlations}

\begin{figure}
\includegraphics[width=8.9cm]{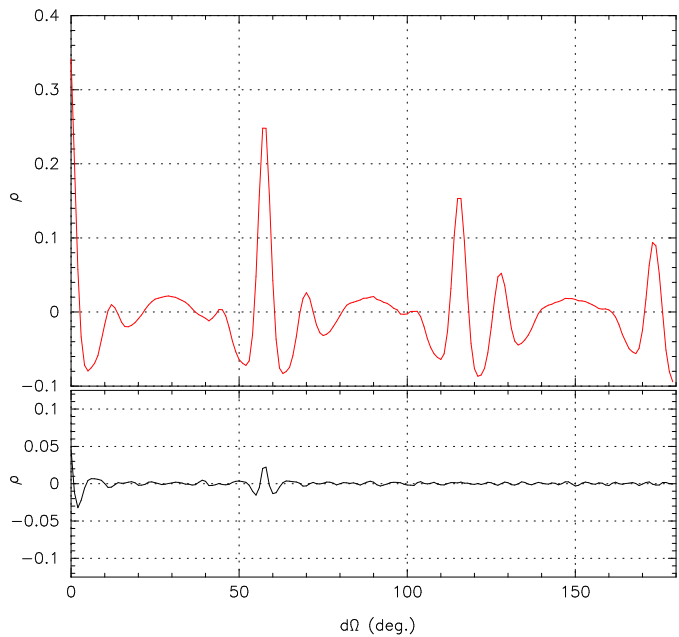}
\caption[]{Abscissa-error correlations for mean orbit transit residuals with 
formal errors below 3 mas as a function of separation along the reference 
great circle. Top: the NDAC data as published in 1997; bottom: the new 
reduction. The peaks are found at intervals of an integer times the basic 
angle of 58~degrees, and fold back at 180 degrees.}
\label{fig:absc_orb_corr}
\end{figure}
A rather troublesome issue with the astrometric catalogue published in 1997
was the correlation level of the abscissa residuals, and the way this affected
the determination of astrometric parameters for, for example, star clusters
\citep[see further][]{vLDWE}. 
Two kinds of determinations of the correlation levels have been 
made for the new reduction, for the field transits and for mean orbit transits.
The latter serves as material for comparison with the 1997 publication, 
as abscissa-error correlations for the published data were determined for mean
orbit transits rather than for the field transits that are available in the 
new reduction.
As abscissa-error correlations originate from inaccuracies in the along-scan 
attitude determination, they naturally show most clearly in the data for the 
brightest transits, where the photon noise is lowest. 
Figure~\ref{fig:absc_orb_corr} shows for the old and the new reduction the
the correlations for mean orbit transits as derived for transits with
formal accuracies better than 3~mas. This essentially selects only the 
brighter stars. The large difference between the NDAC results of 1997 and the 
new reduction is the result of the improved accuracy of the attitude 
modelling. The repeated peaks for the NDAC data show how the great-circle 
reduction process used for building the 1997 catalogue was replicating, up 
to 9 times, local errors in the along-scan attitude solution.

\begin{figure}
\includegraphics[width=8.9cm]{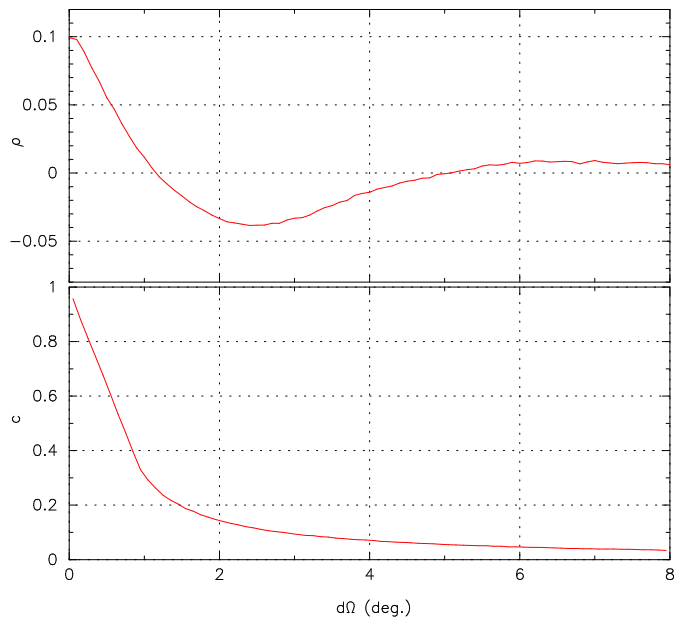}
\caption[]{Top: Abscissa-error correlation level ($\rho$) for field-transit 
abscissa residuals (FTARs) with formal errors below 6 mas, as a function of 
separation along the scan direction. Bottom: Coincidence fraction
$c$ of field-transit measurements as a function of separation on the sky.
The coincidence data are based on a random selection covering 10~per~cent of 
all single stars in the new catalogue.}
\label{fig:absc_frm_corr}
\end{figure}
For the processing of the new catalogue data the correlations between the
FTARs are relevant rather than those between the mean transit residuals per 
orbit. To examine these correlations, data on field transit residuals (with
formal errors below 6~mas) have been collected over all orbits of the mission.
The resulting curve is shown in Fig.~\ref{fig:absc_frm_corr} (upper graph). 
The error limit is about equivalent to the one used above for the mean orbit 
transits.

What is also important for the astrometric data derived from the FTARs is how 
correlations between abscissa residuals can accumulate into correlations 
in astrometric parameters. This is largely determined by the
coincidence factor, as introduced by \citet{fvl99b}. The coincidence factor 
for star $A$ with respect to star $B$ gives the fraction of observations of 
star $A$ for which there are observations of star $B$, contained within the 
same field of view passage. Also taken into account is the typical 
length between nodes in the along-scan attitude modelling, which stretches 
well beyond the width of the field of view. 
The coincidence factor thus defines 
to what extent the underlying abscissa-error correlations could accumulate 
into correlated errors in the astrometric parameters. Coincidence only plays 
a role for stars at relatively short separations on the sky 
(Fig.~\ref{fig:absc_frm_corr}, lower graph). Coincidences with specific stars 
in the other field of view are rarely repeated over the mission. The 
coincidence
level for FTARs is generally about a factor two lower than for mean orbit 
transits, where apparent coincidence would be related to projection on a 
reference great circle rather than actual coincidence of the measurements.

The coincidence factor together with the correlation level determine the 
chance of accumulating correlated errors in the astrometric parameters;
the coincidence fraction defines the average percentage of observations of 
two stars that could be correlated, at a level defined by the correlation 
for that separation (or slightly smaller due to projection effects). 
Considering that the correlation levels in the new reduction are nearly ten 
times smaller than in the 1997 catalogue, and the coincidence factors two 
times smaller, it will be clear that for the new reduction the sensitivity
to accumulation of abscissa-error correlations into astrometric-parameter 
correlations has been very significantly reduced.

\begin{figure}
\includegraphics[width=8.9cm]{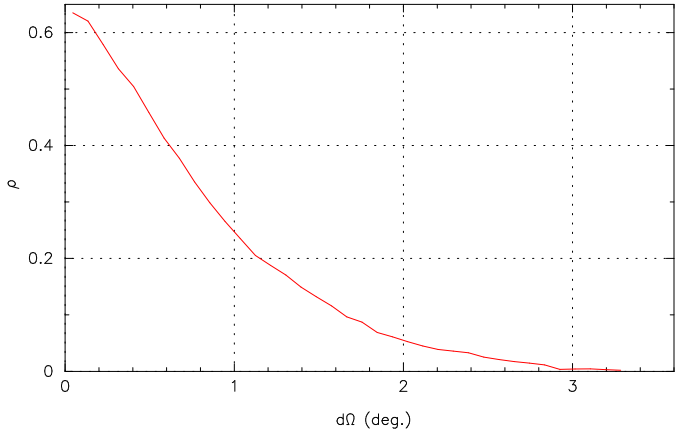}
\caption[]{Parallax correlations in the 1997 catalogue, as observed from
the correlation level $\rho$ between the parallax updates in the new 
catalogue for neighbouring stars as a function of their separation on the 
sky. Only stars with formal errors on the parallax (in the new catalogue) 
below 1.2~mas were selected. Based on the abscissa-error correlations for the
old and the new catalogue, these correlations most likely originate from the
1997 catalogue.}
\label{fig:parr_err_corr}
\end{figure}
Considering the smaller formal errors on the astrometric parameters
in the new reduction, the much lower correlation levels, and the lower
coincidence statistics, the differences between the new and the original
astrometric parameters can reveal correlation levels in the published data.
An example of such test is shown in Fig.~\ref{fig:parr_err_corr} for the
parallax data of stars with formal errors on the parallax less than 1.2~mas
in the new reduction. The level of the correlations is quite high, but their
extent over the sky is small, and agrees, at least qualitatively, with
what could be expected based on the coincidence statistics for mean orbit
abscissa residuals. Beyond a radius of 3 degrees no correlations
are observed, though there may also be (small) anti-correlations existing for
stars at 180 degrees separation, considering the abscissa-error-correlation 
dependencies (Fig.~\ref{fig:absc_orb_corr}) for the 1997 catalogue.

\subsection{Formal errors and their dependencies}

\begin{figure}
\includegraphics[width=8.9cm]{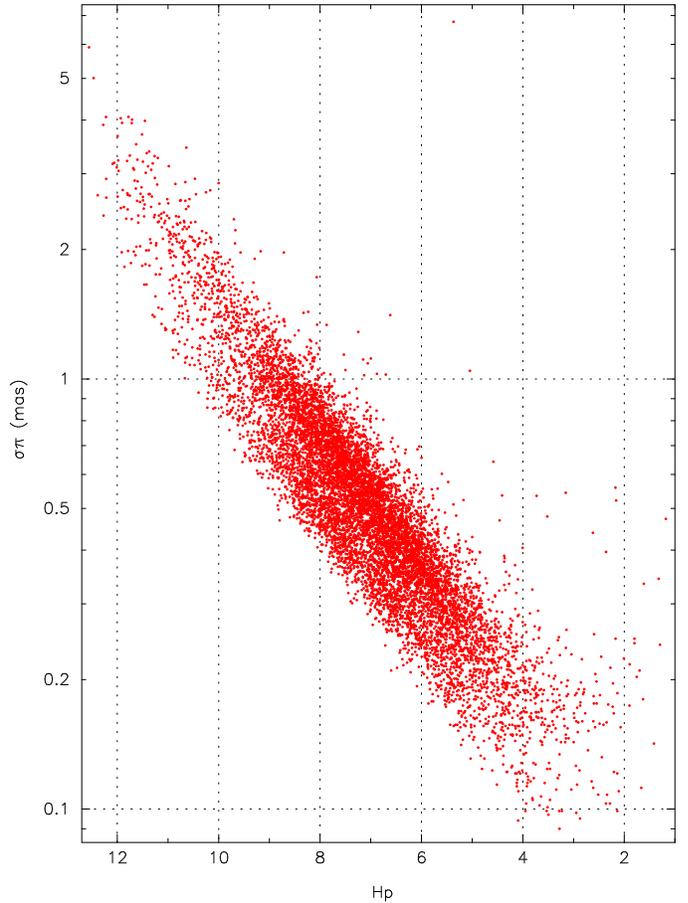}
\caption[]{The formal errors on parallaxes for 9976 stars  with relative errors
on the parallaxes of less than 5~per~cent. The near-linear relation with 
magnitude represents the photon statistics. The two bands in the distribution
represent stars around the ecliptic poles (lower band) and stars around the 
ecliptic plane (upper band).}
\label{fig:hp_parr_err}
\end{figure}
The main error dependencies left in the new catalogue are:
\begin{itemize}
\item The photon noise on the observations, which is a function of the 
magnitude of the star at the time of observation and the (variable) integration
time;
\item The number and distribution of observations, which is a function 
mainly of ecliptic latitude, but which has been locally affected by 
incompleteness in the scan coverage due to lack of ground-station coverage,
observing conditions (like excessively high background signal), and data loss,
most of which was due to hardware problems;
\item Calibration noise, primarily originating from the attitude 
reconstruction.
\end{itemize}
All these contributions can be recognized in Fig.~\ref{fig:hp_parr_err}. The
main linear relation represents the photon statistics. The two discrete
distributions are due to the differences in coverage between the ecliptic poles
(lower band, good coverage) and ecliptic plane (upper band, relatively
poor coverage). The photon noise relations do not continue towards the 
brightest stars, as the errors for these stars are still dominated by 
attitude noise. 

\begin{figure}
\includegraphics[width=8.9cm]{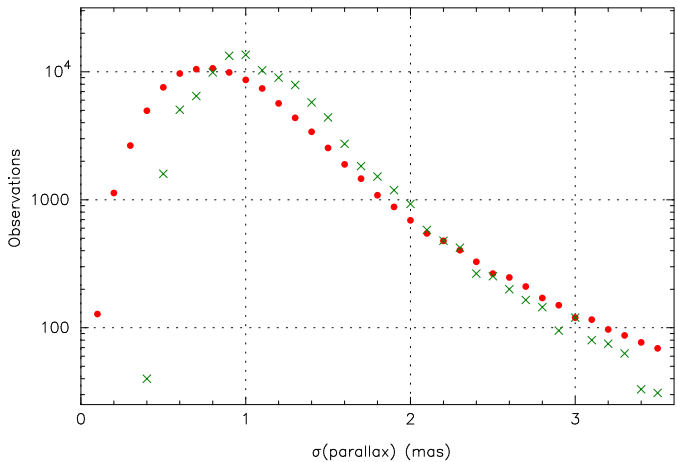}
\caption[]{Histogram of the number of single stars for intervals of width
0.1~mas in parallax error. The dots show the results for the new solution, 
the crosses for the 1997 solution.}
\label{fig:parr_err_hist}
\end{figure}
Figure~\ref{fig:parr_err_hist} shows the distributions of parallax errors 
(below 3.6~mas) for 98~560 stars with simple five-parameter solutions in 
the 1997 publication 
and the new reduction. Some of the discrete features in the distribution
for the 1997 catalogue may be due to the \textit{a posteriori} corrections
that had been applied to formal errors on abscissa residuals.

\subsection{Special cases}

\begin{table}
\caption[]{Published and new determinations for the
astrometry of the primary (HIP~71) and secondary (HIP~70) components of
a wide-binary system.}
\begin{tabular}{llrrrrl}\hline
Star & Param. & Old & $\sigma$ & New & $\sigma$ & units \\\hline
70 & $\varpi$ & $5.25$ & $13.87$ & $5.95$ & $3.08$ & mas \\
71 & $\varpi$ & $9.13$ & $1.84$ & $7.33$ & $1.36$ & mas \\
70 & $\mu_{\alpha *}$ & $-46.78$ & $16.42$ & $-23.25$ & $3.27$ & mas/yr \\
71 & $\mu_{\alpha *}$ & $-24.50$ & $2.05$ & $-23.61$ & $1.39$ & mas/yr \\
70 & $\mu_\delta$ & $-0.88$ & $11.24$ & $-18.58$ & $1.92$ & mas/yr \\
71 & $\mu_\delta$ & $-19.47$ & $1.40$ & $-20.55$ & $0.85$ & mas/yr \\
\hline
\end{tabular}
\label{tab:hip70_71}
\end{table}
For at least two special cases of stars in the Hipparcos catalogue the new
reduction has provided a more than average improvement. The first group
concerns secondary stars in double systems with separate measurements for the
two components, but with the components still sufficiently close to cause
disturbed transits. The improvement on those stars stems from the resolution 
of the data in field transits, which provides much better estimates of the
instantaneous orientation of a double system with respect to the modulating
grid than can be obtained from mean orbit transits. This, in turn, allows for 
a significant improvement in the prediction of the position of the brighter
component of the system relative to the targeted fainter component. An 
accurate relative position with respect to target star and the modulating
grid is essential for correcting the signal of the target star from the
disturbance of the bright companion. An example of how this affected the
astrometric data of such systems is shown in Table~\ref{tab:hip70_71}.
Contrary to the 1997 solution, the new reduction now clearly shows this
system as a physical binary.

The second group of stars are those with extreme red colour indices, many of
which are in addition variable. In the new reduction, epoch-resolved
colour indices have been used, which has two effects. The calibrations
of geometric parameters can also be supported by the very red stars, and
in the application of a calibration, the applicable colour of a star is
more accurately known. Data on the epoch-resolved colours of the red stars
has been provided by Dimitri Pourbaix \citep[see further:][]{platais03,knapp01,
knapp03}.

\subsection{Conclusions on the internal accuracies}

The internal accuracies of the new astrometric data appear to be in all 
aspects consistent with what should be expected on the basis of the
two most important noise contributors, the photon noise and the attitude 
noise. The reduction in correlation level between abscissa residuals is 
in agreement with the reduced noise on the attitude, which has been the 
result of provisions that were made for two major types of disturbance 
(scan-phase discontinuities and hits) in the attitude modelling. 

\section{External accuracy verification}
\label{sec:external}

\subsection{Comparison with radio-star observations}

\begin{figure}
\includegraphics[width=8.9cm]{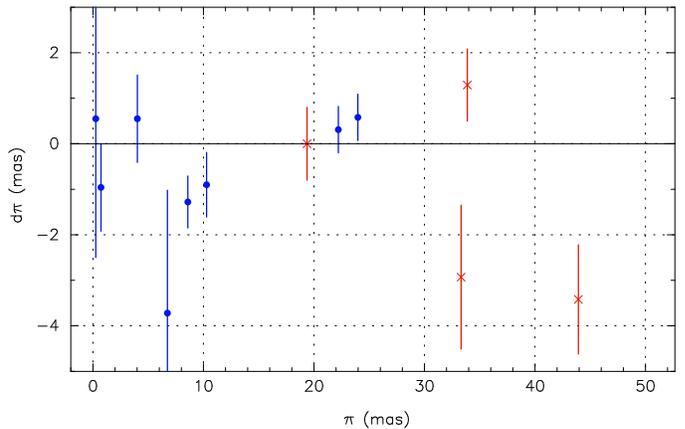}
\caption[]{Differences (VLBI-HIP) in parallax between VLBI measurements 
and the new reduction, as a function of the parallax. Data points indicated
with crosses refer to double stars. Error bars represent the combined 
formal errors on the two sets of observations.}
\label{fig:icrs_par}
\end{figure}
The final transformation of the Hipparcos catalogue to an 
inertial reference frame involved several types of measurements,
and a complete description can be found in Chapter~18 of Volume~3 of
\citet{esa97}. Here the main interest is on whether there are any significant
differences between the 1997 catalogue and the new reduction in as 
far as these transformations are concerned. The focus is on the VLBI
measurements of 12 stars, and the transformations between the astrometric
parameters as determined through radio observations, the 1997 catalogue and
the new reduction.

In the preparation of the linking of the Hipparcos catalogue to the 
ICRS as defined by radio observations \citep{arias95,lestrade95,lestrade99} 
a dozen stars were measured over a period of time also partly covered by the 
Hipparcos mission. Four out of these twelve stars are double stars, and may 
therefore be complicated in their solutions. Some of these stars are in 
addition variable. The radio-star positions have been measured relative to
extra-galactic sources and are thus effectively equivalent to absolute in
parallax and proper motion.  

\begin{figure}
\includegraphics[width=6.5cm]{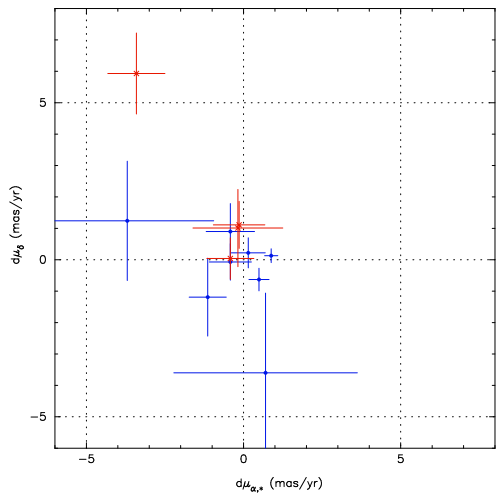}
\caption[]{Differences (VLBI-HIP) in proper motion between VLBI measurements 
and the new reduction. Data points indicated
with crosses refer to double stars. Error bars represent the combined 
formal errors on the two sets of observations.}
\label{fig:icrs_pm}
\end{figure}
\begin{figure}
\includegraphics[width=6.5cm]{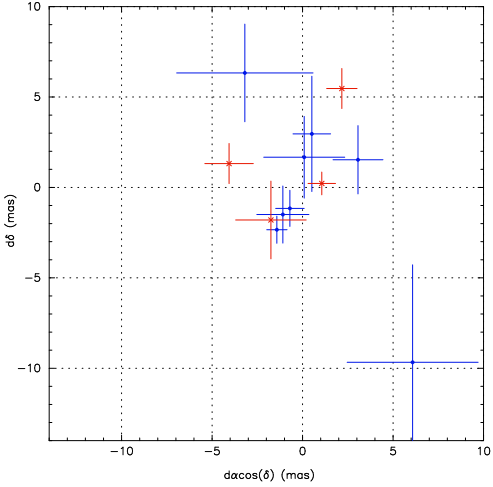}
\caption[]{Differences (VLBI-HIP) in positions between VLBI measurements 
and the new reduction. Data points indicated
with crosses refer to double stars. Error bars represent the combined 
formal errors on the two sets of observations.}
\label{fig:icrs_pos}
\end{figure}
The parallax comparison is shown graphically in Fig.~\ref{fig:icrs_par},
distinguishing between single and double stars. The weighted mean of the
parallax difference for the 8 single stars is determined as 
$-0.013\pm0.48$~mas, with a unit-weight standard deviation (the 
square root of the normalized $\chi^2$) of $1.55\pm0.39$.
A similarly non-significant result is obtained when also including the 
double stars. The VLBI and new Hipparcos parallaxes can thus be considered
to be in good agreement, with a possible hint of an underestimate of the
formal errors in one or both sets of observations.

The transformation in either position or proper motion between two catalogues 
is given by a set of small rotations $(\epsilon_1,\epsilon_2,\epsilon_3)$ 
around three orthogonal axes as:
\begin{equation}
\left[\begin{array}{l}\mathrm{d}\alpha_* \\ \mathrm{d}\delta\end{array}\right]
\approx \left[\begin{array}{rrr}-\cos\alpha\sin\delta & -\sin\alpha\sin\delta &
\cos\delta \\ \sin\alpha & -\cos\alpha & 0 \end{array}\right] \cdot
\left[\begin{array}{l}\epsilon_1 \\ \epsilon_2 \\ \epsilon_3 \end{array}\right].
\label{equ:pos_trans}
\end{equation}
A derivation of these relations can be found in \citet{fvl07}. A nearly
identical set of equations is used for determining the relative spin 
between two catalogues, with on the left-hand side the differences in proper 
motion and on the right-hand side the spin vector \textbf{$\omega$}.
A comparison has been made for the proper motions, and is shown in 
Fig.~\ref{fig:icrs_pm}. The differences in proper motion do not 
represent a significant spin between the two systems. The spin components
are determined as:
\begin{eqnarray}
\omega_1 &=& 0.17\pm0.36~~~\mathrm{mas~yr^{-1}},\nonumber \\
\omega_2 &=& 0.10\pm0.30~~~\mathrm{mas~yr^{-1}},\nonumber \\
\omega_3 &=& 0.40\pm0.34~~~\mathrm{mas~yr^{-1}}, 
\end{eqnarray}
with a unit-weight standard deviation of 1.8 and 12 observations. Excluding
the double stars does not provide an improvement of these results. The
comparison in positions (Fig.~\ref{fig:icrs_pos}) gives a similar result:
\begin{eqnarray}
\epsilon_1 &=& -0.11\pm0.85~~~\mathrm{mas},\nonumber \\
\epsilon_2 &=& \phantom{-}0.06\pm0.86~~~\mathrm{mas},\nonumber \\
\epsilon_3 &=& -0.14\pm0.79~~~\mathrm{mas}, 
\end{eqnarray}
with a unit-weight standard error of 2.0. Given the small number of 
observations, one third of which are double stars, the standard deviations
for these transformations are likely to be affected by peculiarities in the
astrometric parameters of the individual stars, in particular orbital motion.
Differences in the positional origins of the radio and optical signals may also
play a role in a few cases.

\subsection{The optical realization of the ICRS}

The Hipparcos catalogue as published in 1997 serves as the optical realization 
of the International Celestial Reference System or ICRS \citep{feiss98}. This 
section looks at the overall agreement between the new reduction and the 
1997 catalogue in positions and proper motions.

The transformations between the new and old catalogues are again based on 
Eq.~\ref{equ:pos_trans}. The observations are the positional or proper
motion differences between the old and the new catalogue. The errors on 
these observations are partly correlated (when dominated by photon statistics),
depending on the brightness of the object. For the brightest stars there is 
effectively no correlation, 
and the errors are as given by the 1997 catalogue, while for the faintest 
stars the correlation can be considerable. With the brighter stars providing
most of the weight for the solution, errors have been assigned according to
the formal errors given in the published catalogue.
The comparisons are based on the data of 99~130 stars with basic 
five-parameter solutions in both catalogues. In position the following
transformation values are found:
\begin{eqnarray}
\epsilon_x &=& \phantom{-}0.058\pm0.002\quad\mathrm{mas} \\ \nonumber %0.060
\epsilon_y &=& -0.011\pm0.002\quad\mathrm{mas} \\ \nonumber  %-0.010
\epsilon_z &=& \phantom{-}0.028\pm0.003\quad\mathrm{mas}, %-0.030
\end{eqnarray}
with a unit-weight standard deviation of 0.87. As expected, the standard 
deviation is less than 1.0, reflecting the partial correlation of the
data. The same transformation using the formal errors from the new catalogue
for calculating the observation weights gives marginally different results,
and a unit-weight standard deviation of 1.24. In both cases the rotations
observed are, though significant on their own, more than an order of magnitude
smaller than the accuracy with which the positions in the published
catalogue have been linked to the ICRS to provide its optical realization
\citep{arias95,feiss98}.

A similar comparison has been made for the proper motions, giving the
following values:
\begin{eqnarray}
\omega_x &=& -0.001\pm0.003\quad\mathrm{mas/yr} \\ \nonumber %0.001
\omega_y &=& -0.005\pm0.003\quad\mathrm{mas/yr} \\ \nonumber %-0.014
\omega_z &=& +0.006\pm0.003\quad\mathrm{mas/yr},  %0.004
\end{eqnarray}
and a unit-weight standard deviation of 0.90 (or 1.26 when new weights are
applied). It can therefore be concluded that the new catalogue as a reference
frame is essentially identical to the 1997 catalogue, and only very small 
and insignificant differential rotations have been introduced in its 
preparation. 

\subsection{Parallaxes}

\begin{figure}
\includegraphics[width=8.9cm]{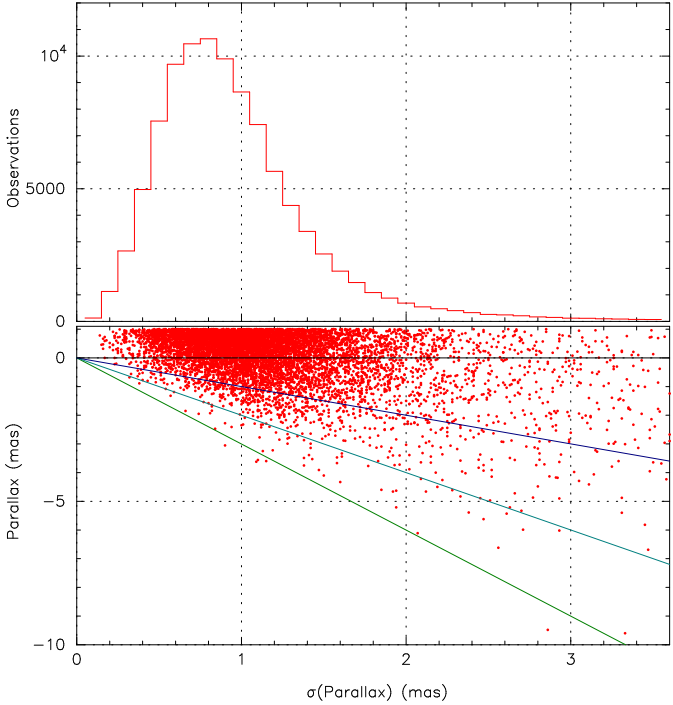}
\caption[]{Top: the histogram, for all single stars with five-parameter 
solutions, of the formal error on the parallaxes. Bottom: the distribution of 
parallaxes less than $1~\mathrm{mas~s}^{-1}$ as a function of
formal error on the parallax determination for the new solution. The 
diagonal lines show the one, two and three sigma levels as based on the 
formal errors. }
\label{fig:negat_par}
\end{figure}
The verification of formal errors on parallaxes in the published 
catalogue relies primarily on the 
observed distribution of negative parallaxes. Such verification is only
partially possible with the new reduction. The main reason is that amongst
those stars for which a negative parallax was derived, the smallest
formal error is 0.23 mas. There are no negative parallaxes amongst the 
1000 stars with smaller errors than this value, and only two for the first 
2400 stars, when ranked according to formal parallax error. 
Figure~\ref{fig:negat_par} shows the
overall distribution of negative parallaxes as a function of the formal
errors, as well as the histogram of the numbers of stars in intervals of 
0.1~mas width in formal error. 

\begin{figure}
\includegraphics[width=8.9cm]{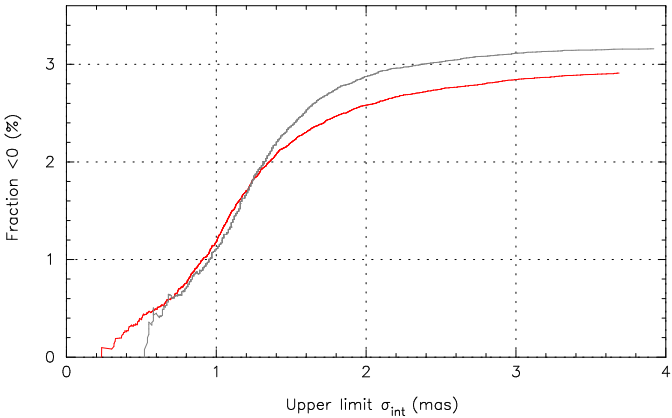}
\caption[]{The cumulative percentage of negative parallaxes as a function
of the formal error on the parallax, $\sigma_\mathrm{int}$. The grey curve
in the background shows the same information as derived from the 1997
catalogue.}
\label{fig:cumperc_par}
\end{figure}
Considering the discussion of the internal errors in the preceding section
it seems incorrect to apply any form of error scaling as was done for
the 1997 catalogue \citep[see Volume 3 of ][]{esa97}. What could be considered
is a random background noise that can not be detected in the internal error
analysis. This could be noise left in the reference frame as determined 
through the iterations between the along-scan attitude reconstruction and 
the astrometric-parameter determination. This would most likely be a 
residual effect still left over from the 1997 catalogue. Such noise residual 
could be noticeable for those stars with the smallest formal errors.  
However, among those there is only a very low number of stars with negative 
parallaxes, far too few for any objective statistical treatment, as can
be seen from Fig.~\ref{fig:negat_par} and Fig.~\ref{fig:cumperc_par}. The
overall decrease in the total number of negative parallaxes is one of the
clear indications for the improvements achieved with the construction of
the new catalogue.

\begin{table}
\caption[]{Data on the ten stars with negative parallaxes and the smallest
formal errors.}
\scriptsize{
\begin{tabular}{rrrrrrll}\hline
HIP & $\pi$ & $\sigma_{\pi}$ & Hp & $\mathrm{B-V}$ & HD & Name & Spectr.\\
\hline
103312 & $-0.22$ & 0.23 & 5.801 &  0.408 & 199478 & V2140~Cyg  & B8Ia  \\       
 57741 & $-0.20$ & 0.24 & 5.761 &  0.233 & 102878 &            & A3Iab \\      
 54751 & $-0.03$ & 0.30 & 4.718 &  0.541 &  97534 & V533~Car   & A6Ia  \\      
 44904 & $-0.68$ & 0.32 & 6.861 &  0.150 &  78949 &            & A1/A2III \\   
107749 & $-0.21$ & 0.32 & 6.569 &  0.381 & 207673 &            & A2Ib    \\    
 61703 & $-0.56$ & 0.33 & 6.269 &  0.007 & 109867 & KY~Mus     & B1Ia    \\    
 52004 & $-0.16$ & 0.33 & 5.551 &  0.500 &  92207 & V370~Car   & A0Ia    \\    
 97485 & $-0.01$ & 0.34 & 6.490 &  0.129 & 187459 & V1765~Cyg  & B0.5Ibvar \\  
 88298 & $-0.13$ & 0.36 & 5.732 & -0.030 & 164402 &            & B0Iab...  \\  
 41074 & $-0.27$ & 0.37 & 5.979 &  0.379 &  70761 &            & F2Iab     \\  
\hline
\end{tabular}
}
\label{tab:negatpar}
\end{table}
Given the very low numbers of negative parallaxes for those stars with the
smallest formal errors, it seems more appropriate to examine the individual
cases for which a negative parallax has been obtained rather than to treat 
these as some statistical variable. The ten stars with negative parallaxes and
the smallest formal errors are presented in Table~\ref{tab:negatpar}.
All but one of these stars are supergiants. The spectral information for
star HIP~44904 was given by \citet{eggen86}, who stated that the data 
for this star could also be interpreted as coming from a high-mass 
supergiant. The latter seems to be the case, judging by the parallax 
information. It is also clear from a comparison between colour indices and
spectral types that most of these stars are considerably reddened, which
should not be surprising given that for most the distance is well over 1~kpc.

\begin{figure}
\includegraphics[width=8.9cm]{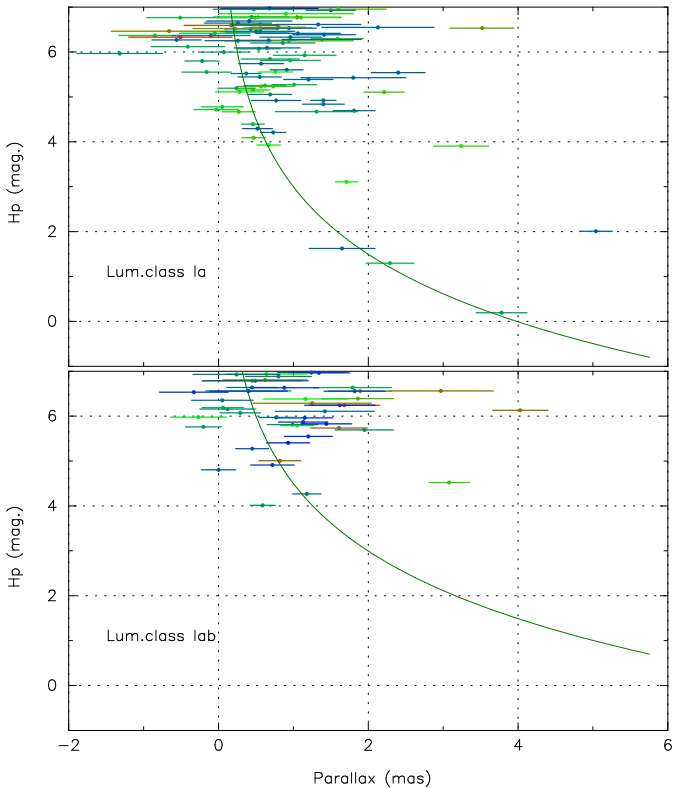}
\caption[]{The distribution of single stars with luminosity class Ia (top)
and Iab (bottom) in measured parallax and Hp magnitude. The magnitudes
have not been corrected for reddening. The curve in the upper diagram
represents an absolute magnitude of M$_\mathrm{V}=-7.0$, and in the lower 
diagram for M$_\mathrm{V}=-5.5$. The observed spread is due to both the 
formal errors on the parallaxes (as shown) and the (unspecified) 
formal errors on the spectral classifications.}
\label{fig:lumclassiab}
\end{figure}
Unfortunately, there is considerable uncertainty (or intrinsic spread) in 
the absolute magnitudes of these supergiants \citep[see also][]{wegner07}
and the reddening corrections to be applied to the observed magnitudes
\citep{wegner03}. Also the spectroscopic classification is not always
uniquely defined. From an examination of the distribution in parallaxes 
and apparent magnitudes of the type Ia stars, it seems that the brightest of 
these supergiants have an absolute magnitude of around $-7$, 
but there could be many fainter by up to three magnitudes (though this is
at least partly caused by reddening). There is an indication of some stars
still being brighter. These could be the so-called ``hypergiants'', with
absolute magnitudes of around $-8$. An example is the star HIP~95657
(HD~183143), identified as a possible hypergiant at a distance of 2~kp by
\citet{chentsov04}, for which in the new reduction a parallax of $0.47\pm0.54$ 
has been determined.
There are no type Iab stars close enough for a comparable estimate. These
results are illustrated by Fig.~\ref{fig:lumclassiab}, showing all supergiants
with apparent magnitude brighter than $\mathrm{Hp}=7$.
Given the uncertainties in classification as well as associated absolute
magnitudes and reddening corrections, no firm conclusions should be drawn 
from the observed distribution of negative parallaxes.

There is, however, another approach that can be made. The distribution
of negative parallaxes and their formal errors can be represented as a 
unit-weight error distribution, assuming all actual parallaxes are zero as
an absolute lower limit. In that case the negative parallaxes, divided by
their formal errors, represent one half of a Gaussian distribution with
$\sigma=1$. The $\chi^2$ of the parallax measurements for the
ten stars in Table~\ref{tab:negatpar} equals 10.34, with 9 degrees of freedom. 
The P-value is 0.32, so the test gives no ground to reject the assumption
that the formal errors are correct. Such 
procedure can in principle also be extended to a non-zero parallax
assumption, but that would have to be based on assumed absolute magnitudes,
and would require reddening corrections and the inclusion of all stars
with parallaxes smaller than predicted, and not only those with negative 
parallaxes. This would then make the calculations model dependent. A simplified
experiment shows, however, that even if we assume all reference parallaxes
to equal 0.2~mas, the P-value of the $\chi^2$ test would be as high as 0.17,
thus still not inconsistent with the assumptions.
 
Concluding, it seems from the data that could be analyzed that there is no
proof for the presence of an additional noise contribution to the parallax
errors. However, the detection capability of the data for such contribution, 
at a level of a few tenths of a mas, is not very high, and its presence 
can not be excluded either. Findings for the internal accuracies and the
convergence of the new reduction make such noise contributions not very 
likely. 

\section{The potential impact of the new reduction}
\label{sec:impact}

\begin{table}
\caption[]{Comparisons of accumulated weight statistics for parallax
measurements of single stars with five-parameter solutions in the old and 
the new Hipparcos reductions. The errors are given in mas.}
%\scriptsize{
\begin{tabular}{rrrrrrll}\hline
& \multicolumn{2}{c}{Observations} & \multicolumn{2}{c}{Mean error} & Weight\\
Interval & New & Old & New & Old & ratio\\
\hline
All & 101783 & 100037 & 0.66 & 0.96 & 2.16 \\
$\mathrm{Hp}\ge 9$ & 34799 & 33347 & 1.23 & 1.33 & 1.23 \\
$\mathrm{Hp}<9$ & 66984 & 66690 & 0.56 & 0.86 & 2.36 \\
$\mathrm{Hp}<7$ & 10883 & 11127 & 0.33 & 0.70 & 4.29 \\ 
\hline
\end{tabular}
%}
\label{tab:weights}
\end{table}
The potential impact of the new reduction can be estimated from the 
formal errors on for example the parallaxes of the single stars. Considering
only simple five-parameter solutions, the combined weight for a selection of 
stars in either the new or the old solution is given by the sum of the
inverse squared formal errors:
\begin{equation}
w=\sum_i (1/\sigma_\pi)^2.
\end{equation} 
The change in weight is then given by the ratio of the $w$ values for the
new and old reductions. For the entire catalogue, the increase in weight 
achieved by the new reduction equals a factor 2.16. Details for 
different magnitude ranges are presented in Table~\ref{tab:weights}.
A decrease is noted for the number of bright stars with five-parameter
solutions in the new reduction. This is the result of the increased 
sensitivity at higher 
accuracy to any orbital disturbance. Somewhat of a surprise has been the 
improvements observed for the faint stars. However, already the 
comparisons between the FAST and NDAC data in the 1997 reduction showed
that the noise on the data for these stars as published in the old catalogue 
also contained a significant instrument-modelling component, as was shown by
the error correlation statistics \citep[see Volume 3 of ][]{esa97}.

The largest impact of the new reduction is likely to be for Cepheids and
for star clusters and associations. For the Cepheids we examine stars of 
relatively high intrinsic brightness and at distances above 300~pc. Using the
selection criteria for the Cepheid study of \citet{feastcatch97}, about
four times more stars would be selected. This shows in a study based on 
the one-but-final iteration results obtained in the construction of 
the new catalogue by \citet{fvlfeast07}. While based on the 1997 reduction
only the zero point of the Cepheids PL relation could be determined
\citep{feastcatch97}, there is
now also significant information available on the slope of this relation
\citep[see also ][]{fvl07}.

For open cluster studies preliminary results are presented by \citet{fvl07},
which show an improvement of the formal errors by about a factor 2 to 2.5,
with formal errors on the parallaxes of the 8 nearest clusters in the range 
of 0.09 to 0.19~mas. The gains for the open clusters are made in three areas:
\begin{itemize}
\item The improvements in the formal errors for the brightest stars;
\item The significant reduction in the correlations between the abscissa 
errors;
\item The much improved handling of the connectivity condition in the 
along-scan attitude reconstruction.
\end{itemize} 
The second point simplifies the reduction of the open cluster data to
single cluster parallax and proper motion values considerably. The third
point avoids the potential Pleiades problem to which open cluster data 
in particular can be very sensitive. This is due to the concentrations of
relatively bright stars and the weight these may potentially assert in 
the along-scan attitude reconstruction. The open cluster analysis and results 
will be presented in a separate paper, currently in preparation.

\begin{figure}
\includegraphics[width=8.9cm]{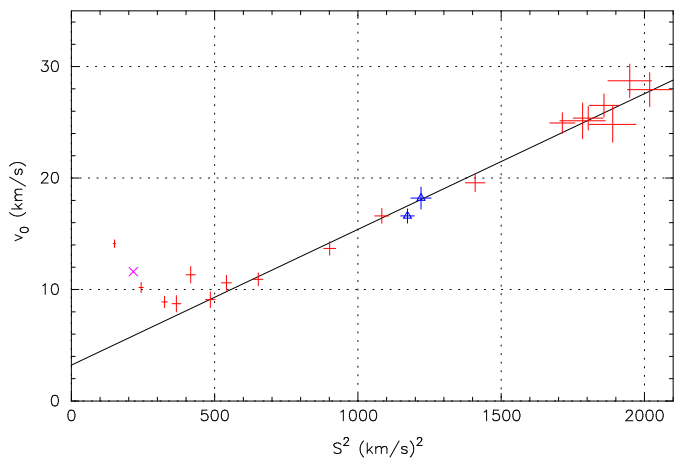}
\caption[]
{The asymmetric drift in the galactic rotation shown as the relation between
the observed values of $v_0$ and $S^2\equiv 2\sigma^2$ for main sequence
stars and Red Clump and RGB stars (indicated by triangles). The value
observed for the Cepheids is shown by $\times$ \citep[from ][]{fvl07}.}
\label{fig:solmotdrift}
\end{figure}
A significant impact can also be made on studies of galactic dynamics. 
A repeat of the study by \citet{binney00}, for example, shows a factor
two more stars available, resulting in additional measurements towards 
later spectral types \citep{fvl07}. Three more data points could be added 
to the asymmetric drift diagram, for Cepheids, Red Clump and RGB stars,
as is shown in Fig.~\ref{fig:solmotdrift}. Preliminary results on
the solar motion and galactic rotation, as based on the new reduction, can 
also be found in \citet{fvl07}.

%\subsection{Proper motions}
\section{Conclusions}
\label{sec:conclusion}

The possibilities to verify astrometric data down to a level of 
0.1~mas are rather limited, in particular for relatively bright stars. 
The internal verification has to provide the main
body of evidence, as almost no data is available for an independent external 
verification. Independent here means free from assumptions derived from the
application of those data to the analysis of astrophysical objects, and in 
particular when that involves assumed absolute magnitudes and their intrinsic
dispersion. The internal verification of the new reduction appears to provide
fully self-consistent results and, as such, provides the main element of the
data quality verification. The limitations of the external verification
leave open a number of questions, for which the answers may not be 
available until a much improved independent catalogue, such as the one 
expected from the Gaia mission, is available. 

The improvements achieved by the new reduction of the Hipparcos astrometric 
data put the results of this mission significantly above its original aim, 
which was set at 2~mas. In setting this aim, the efficiency of the detectors 
had been underestimated, and the noise contribution of the instrument 
calibrations overestimated. The original reductions, as published in 1997, 
already provided data that was a factor two better on average. The new 
reduction has made this a factor three on average, and more than a factor ten
for many of the brightest stars. With these improvements, significant 
progress can again be made in a wide range of luminosity calibration studies 
and studies of galactic dynamics.

Above all, however, the new reduction has shown that the principle of Hipparcos
for obtaining absolute parallaxes works, though it requires very careful
implementation of various aspects of the data processing, and in particular
the along-scan attitude reconstruction. The latter could only be achieved
for the Hipparcos mission through a thorough understanding of the satellite
dynamics.

\begin{acknowledgements}

I would like to express my thanks to Lennart Lindegren, Dafydd W.~Evans, 
Rudolf Le Poole and Elena Fantino for their support,  suggestions, ideas and 
corrections along the long way of this study.
 
\end{acknowledgements}

\bibliographystyle{aa}
\bibliography{8357}
\end{document}